\documentclass[twocolumn,english,prl]{revtex4-1}
\usepackage[T1]{fontenc}
\usepackage[latin9]{inputenc}
\usepackage{amstext}
\usepackage{amssymb}
\usepackage{graphicx}
\usepackage{babel}
\begin{document}

\title{Percolation transition in the packing of bidispersed particles on
curved surfaces}

\author{Andrew M. Mascioli}

\affiliation{Department of Physics and Astronomy, Tufts University, 574 Boston
Avenue, Medford, Massachusetts 02155, USA}

\author{Christopher J. Burke}

\affiliation{Department of Physics and Astronomy, Tufts University, 574 Boston
Avenue, Medford, Massachusetts 02155, USA}

\author{Timothy J. Atherton}
\email{timothy.atherton@tufts.edu}

\selectlanguage{english}%

\affiliation{Department of Physics and Astronomy, Tufts University, 574 Boston
Avenue, Medford, Massachusetts 02155, USA}
\begin{abstract}
We study packings of bidispersed spherical particles on a spherical
surface. The presence of curvature necessitates defects even for monodispersed
particles; bidispersity either leads to a more disordered packing
for nearly equal radii, or a higher fill fraction when the smaller
particles are accomodated in the interstices of the larger spheres.
Variation in the packing fraction is explained by a percolation transition,
as chains of defects or scars previously discovered in the monodispersed
case grow and eventually disconnect the neighbor graph. 
\end{abstract}
\maketitle
Bidispersed mixtures of hard spheres are an important elementary model
of a glass transition\cite{stillinger2013glass}: at high temperature
and low density they flow freely, while as temperature is reduced
they become kinetically arrested and form rigid but highly disordered
structures\cite{Donev2004}. At zero temperature and stress, a similar
\emph{jamming} transition to rigidity occurs as a function of density\cite{cates1998jamming,Liu2010}
which in 2D tends to occur around a packing fraction of $\Phi=0.84$
\cite{OHern2002,reichhardt2014aspects}. Jammed structures exhibit
distinctive properties including \emph{isostaticity}: the average
number of inter-particle contacts is the minimum number required for
mechanical stability\cite{Alexander1998}. Powerful mathematical tools
exist\cite{Donev2004b} to classify jammed and glassy packings of
hard particles according to a hierarchy, depending on where individual
particles, groups or boundary deformations can unjam the system\cite{Torquato2001}. 

Sphere packings, the high density and zero temperature limit of these
processes, have been extensively studied in both 2D and 3D Euclidean
space\cite{Liu2010,Torquato2010a,Donev2004,Lubachevsky1991} revealing
strong dimensional dependence: 2D monodispersed spheres tend to crystallize
readily, because the locally dense hexagonal packing fills space;
in 3D the locally dense tetrahedral packing cannot fill space, permitting
a random close packed structure that is the subject of much debate\cite{torquato2000random,donevcomment,kamien2007random}.
Even in 2D, however, disorder can be induced in bidispersed systems.
Molecular dynamics simulations have shown that there is a transition
from order to disorder as the degree of bidispersity is increased\cite{Hamanaka2006,Sadr-Lahijany1997,Vermohlen1995,Watanabe2005},
and statistical models of bidispersed particle packings have been
used to predict the local features of disordered bidispersed packings\cite{Hilgenfeldt2013,Richard2001}.
The degree of order or disorder can be measured by an order parameter
such as the hexatic bond orientational order\cite{Nelson1979}.

Crystalline order is geometrically frustrated on curved surfaces\cite{Seung1988}:
an incompatibility between the preferred hexagonal symmetry of the
crystalline packing and the topology of the surface necessitates a
minimal number of defects\textemdash particles with a number of neighbors
other than 6\textemdash to accommodate the curvature. For monodispersed
particles, the packings are mainly crystalline with a transition between
isolated defects for small particle number and chains of defects or
\emph{scars} akin to grain boundaries in bulk systems that occur above
a critical number of particles $N_{c}\approx110$ and grow with system
size\cite{Bausch2003,Bowick2000}. The scars may join in asterisk-like
motifs\cite{Bowick2000} and are aligned by anisotropic curvature\cite{Burke2015}.
Jammed packings on spheres or \emph{spherical codes }have recently
been studied in multiple dimensions \cite{Cohn2011}.

In this Letter, we investigate the packing of bidispersed particles
on a spherical surface as a simple model of how glasses interact with
curvature. We determine the packing fraction, connectivity and hexatic
order parameter as a function of particle number $N$, fraction of
large particles $\chi=N_{L}/N$ and bidispersity $b=\left(r_{1}-r_{2}\right)/\left(r_{1}+r_{2}\right)$
where $r_{1}$ and $r_{2}$ are the radii of the particles and $r_{1}\ge r_{2}$.
By identifying topological defects from the neighbor graph we show
that variation in these parameters is explained by a percolation transition
due to growth and connectivity of the scar network, as well as by
the possibility of commensurate local packings. 

\emph{Simulations\textemdash }Packings with high coverage fraction
were produced using a surface relaxation algorithm: $N$ spherical
particles are initially placed using random sequential absorption
with their centers of mass on a sphere of radius $R=1$. Particles
are randomly assigned to two categories corresponding to larger and
smaller radii respectively. The simulation proceeds by, first, \emph{diffusion
sweeps} where, particles are moved in random order some distance drawn
from a Gaussian distribution of width $\sigma=2r_{1}\times10^{-3}$
in a random direction along the surface. Moves that cause overlap
are rejected. As the packing becomes dense, an adaptive step size
is used to reduce the number of moves rejected due to overlap: $\sigma=10\langle s\rangle$,
where $\langle s\rangle$ is the geometric mean of the separation
between each particle and its three nearest neighbors. Secondly, \emph{surface
relaxation} moves slowly decrease the radius of the surface by an
amount $\Delta R$, where initially $\Delta R=10^{-5}$. After the
surface radius is reduced, particles are projected down onto the nearest
point on the surface. After projection, a gradient descent minimization
is run on the particles (where the interparticle energy is linear
in the amount of overlap) until overlap is undone. If overlap can
not be undone, the surface relaxation move is undone and particle
positions are reset, and simulation continues with $\Delta R$ set
to $\Delta R/2$. $20$ diffusion sweeps are carried out between each
surface relaxation step. The simulation halts when $\text{\ensuremath{\Delta}R}$
is reduced to $2^{-14}$ times its original value. 

\begin{figure}
\includegraphics[width=1\columnwidth]{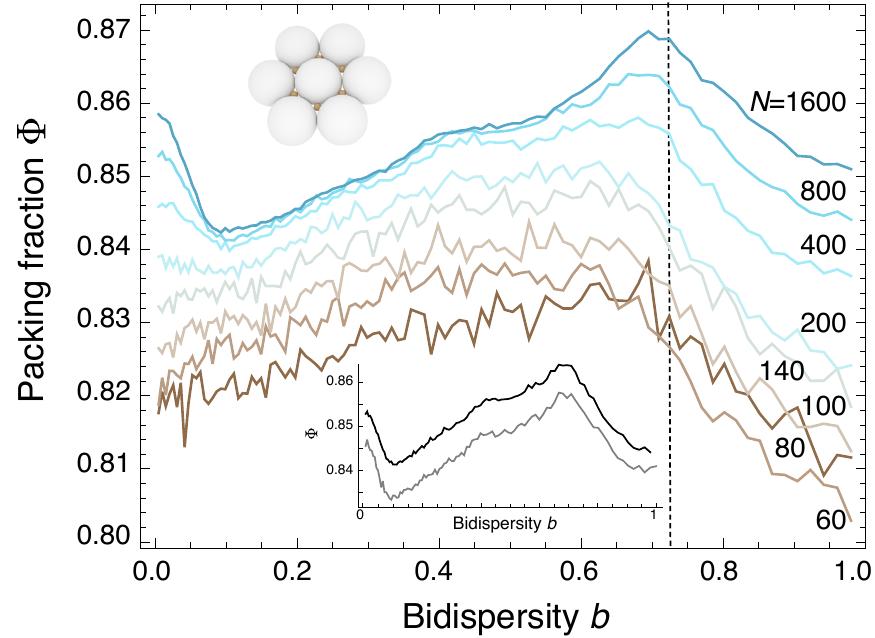}

\caption{\label{fig:Packing}(Color online) Packing fraction $\Phi$ as a function
of bidispersity $b=\left(r_{1}-r_{2}\right)/\left(r_{1}+r_{2}\right)$
where $r_{1}>r_{2}$ for different particle numbers $N$. The maximum
$b=\sqrt{3}-1\approx0.73$, indicated with a vertical dashed line,
occurs for an Apollonian packing, i.e. where smaller particles fit
in the interstices of the larger particles as depicted in the upper
inset. \emph{Lower} \emph{inset:} comparison of the packing fraction
of arrested (gray) and jammed (black) packings for $N=800$ particles.
}
 
\end{figure}

Configurations produced by this procedure are referred to as \emph{arrested},
because they remain metastable if the simulation is restarted; eventually,
however, a Monte Carlo move will unjam the arrested configuration,
potentially facilitating further relaxation and a consequent increase
in the packing fraction. This process occurs in real glasses and is
known as \emph{aging}. Extending a powerful technique due to Donev
\emph{et al.} \cite{Donev2004b}, we artificially age the arrested
structures using a linear program to find and execute an \emph{unjamming}
motion of the particles and further relax the surface. Iterative unjamming
and relaxation guides the packing toward a state that is collectively
jammed with respect to movement of the particles and further relaxation.
As we report elsewhere\cite{atherton16}, the convergence of this
procedure is greatly accelerated by preconditioning the packing, attaching
a short range repulsive interaction to the particles beyond the hard
inter-penetrability constraint and minimizing the corresponding energy
by gradient descent. This procedure moves the particles into the center
of the feasible region from which the linear program is more effectively
able to identify an unjamming motion. Each arrested structure was
subjected to this artificial aging process to produce a corresponding
ensemble of \emph{jammed} structures. 

For monodispersed particles\cite{Bausch2003}, neighbors are assigned
from a Voronoi tessellation\cite{aurenhammer1991voronoi} of the particle
centers of mass, partitioning the surface into $N$ polygonal regions
closest to a particular particle. Two particles are neighbors if they
share an adjacent edge on the Voronoi tessellation. Generalizing this
construction to bidispersed particles with a weighted distance fails
to uniquely assign all points on the surface to a particle; two proposed
alternatives \cite{Richard2001} are the r\emph{adical tessellation}
and the \emph{navigation map}, both of which recover the Voronoi tessellation
in the limit of monodispersed spheres. The radical tessellation utilizes
the radical plane as a separatrix between each pair of particles;
the navigation map partitions the surface into regions closest to
the surface of the particles rather than their center of mass. We
found little difference between quantities calculated from these constructions
and use the radical tessellation exclusively in the remainder of the
paper. From the radical tessellation, the adjoint neighbor graph was
constructed for each packing and the coordination number determined
for each particle. 

\emph{Results and Discussion\textemdash }For each value of bidispersity
on the interval $b\in[0,1]$ with a resolution of $\Delta b=0.005$,
an ensemble of 20 jammed configurations was generated with $\chi=1/2$
and for different numbers of particles $N$. The packing fraction
$\Phi$, i.e. the fraction of the surface enclosed by the particles,
was calculated for each configuration and shown in fig. \ref{fig:Packing}.
For particle numbers above about $N=200$, slight deviations from
the monodispersed case immediately introduce disorder and reduce the
packing fraction as expected. Above a critical value of bidispersity
$b_{c}\sim0.1$, however, we see a transition and $\Phi$ increases,
with an apparent shoulder at $b\approx0.4$, up to a maximum value
of $\Phi\approx0.87$ at $b=b_{A}\sim0.7$ and then decreases as $b\to1$.
For $N<200$, $\Phi$ increases monotonically up to a maximum at a
slightly lower value of $b\sim0.6$. In the lower inset of Fig. \ref{fig:Packing},
we compare the packing fraction for 800 particles for the ensemble
of arrested and jammed packings. It is clear that the arrested structures
are slightly less efficiently packed, but the trends are identical.
We find simular results for all $N$; this correspondence affirms
that the trends are geometric in origin rather than due to variation
in the performance of the algorithm at different $b$. 

\begin{figure}
\includegraphics[width=1\columnwidth]{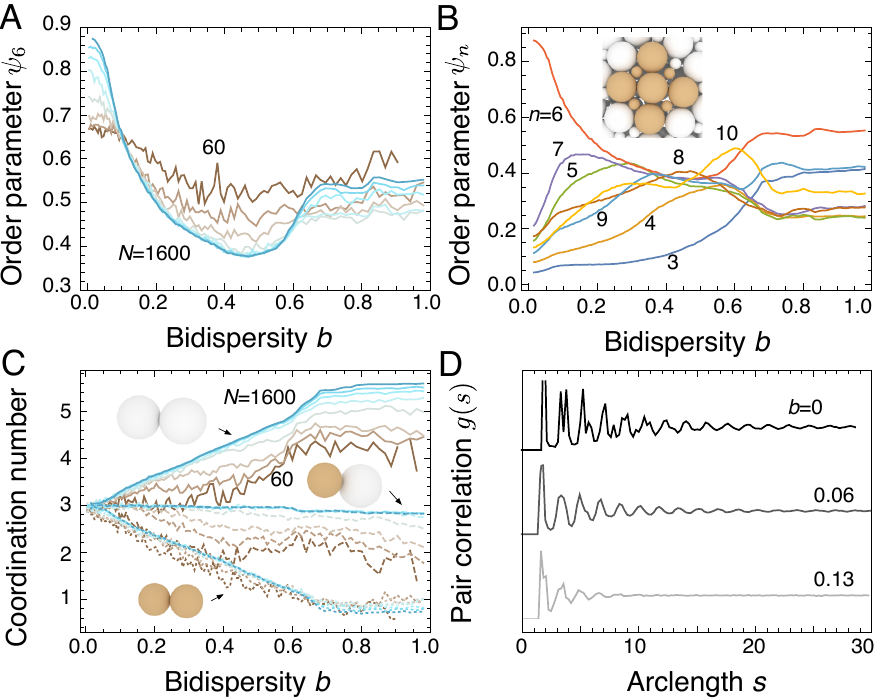}

\caption{\label{fig:OrderParameter}(Color online) \textbf{A }Hexatic order
parameter as a function of bidispersity $b$ for different $N$. \textbf{B}
Local order parameters $\psi_{n}$ as a function $b$ for a jammed
packing of $N=1600$ particles. Hexatic order dominates except intermediate
values of $b$ where eight- and ten-fold order possess maxima. \emph{Inset}:
Commensurate configuration with high coordination number at $b=\sqrt{2}-1\approx0.41$.
\textbf{C }Average coordination number for large-large (solid lines),
large-small (dashed lines) and small-small (dotted lines) inter-particle
contacts for varying $N$. \textbf{D }Pair correlation function $g(s)$
for three values of bidispersity. }
 
\end{figure}

The maximum at $b=b_{A}$ is immediately explicable: it corresponds
to the special point at which the smaller particles fit exactly in
the interstices between the larger particles, depicted in the upper
inset of Fig. \ref{fig:Packing}. We denote this the \emph{Apollonian
point} in reference to the tiling. Packings around and above $b_{A}$
appear mostly crystalline with the smaller particles separated into
the interstices; the packing fraction at $b=1$ corresponds exactly
to that for $N/2$ particles. No such immediate explanation is obvious
for the low and medium bidispersity results, which appear to be well
mixed; we therefore seek a more detailed understanding of the structure. 

One structural measure that reflects the degree to which the packings
are locally crystalline is the hexatic order parameter $\psi_{6}=\left\langle \exp(i6\theta_{i})\right\rangle $,
where the average is taken over the neighboring particles. This is
shown calculated from the dataset as a function of $b$ and $N$ in
Fig. \ref{fig:OrderParameter}A. A maximum occurs for all $N$ at
$b=0$ as expected; the value is reduced for smaller $N$ reflecting
the disruption of crystallinity by the curvature. The hexatic order
drops with $b$, reaches a minimum around $b\sim0.45$, rises and
then forms a plateau above the Apollonian point, albeit at a value
significantly lower than the $b=0$ case, because here the large particles
have a higher coordination number. Variation in $\psi_{6}$ is significantly
attenuated for low $N$ where the influence of the curvature is stronger. 

To see whether hexatic order is replaced by other ordering, we calculated
$n$-atic order parameters $\psi_{n}=\left\langle \exp(in\theta_{i})\right\rangle $
for $N=1600$ as a function of $b$; the results are plotted in Fig.
\ref{fig:OrderParameter}B. In contrast to the hexatic order parameter,
$\psi_{n}$ for $n\neq6$ \emph{increases} with $b$ from $b=0$;
moreover all $\psi_{n}$ exhibit a plateau above the Appollonian point
confirming the distinct nature of this regime. Two values, $n=8,10$
have $\psi_{n}$ narrowly greater than $\psi_{6}$ for intermediate
values of $b$ and possess maxima at $b=0.45$ and $b=0.6$ respectively.
Examining the packings, this is due to the presence of octagonally
and decagonally coordinated arrangements: a common and commensurate
motif, depicted in the inset of fig. \ref{fig:OrderParameter}B, where
four large and four small particles are arranged around a central
large particle, is allowed first for $b=\sqrt{2}-1\approx0.41$, which
coincides with the position of the shoulder in the plot of $\Phi(b)$
in Fig. \ref{fig:Packing}. A variety of similar motifs exist for
$b$ around this value with the same coordination number but different
mixtures of large and small neighboring particles and appear to cause
the shoulder. It is interesting to note significant decatic ordering:
10-fold rotational symmetry is incompatible with long range order
and is rarely seen in packings in flat space with the exception of
quasicrystals\cite{PhysRevLett.53.1951,Fischer01022011,talapin2009quasicrystalline}.
As long range order is also incompatible with curvature, it appears
that curvature may promote the increased 10-fold ordering.

\begin{figure}
\includegraphics[width=1\columnwidth]{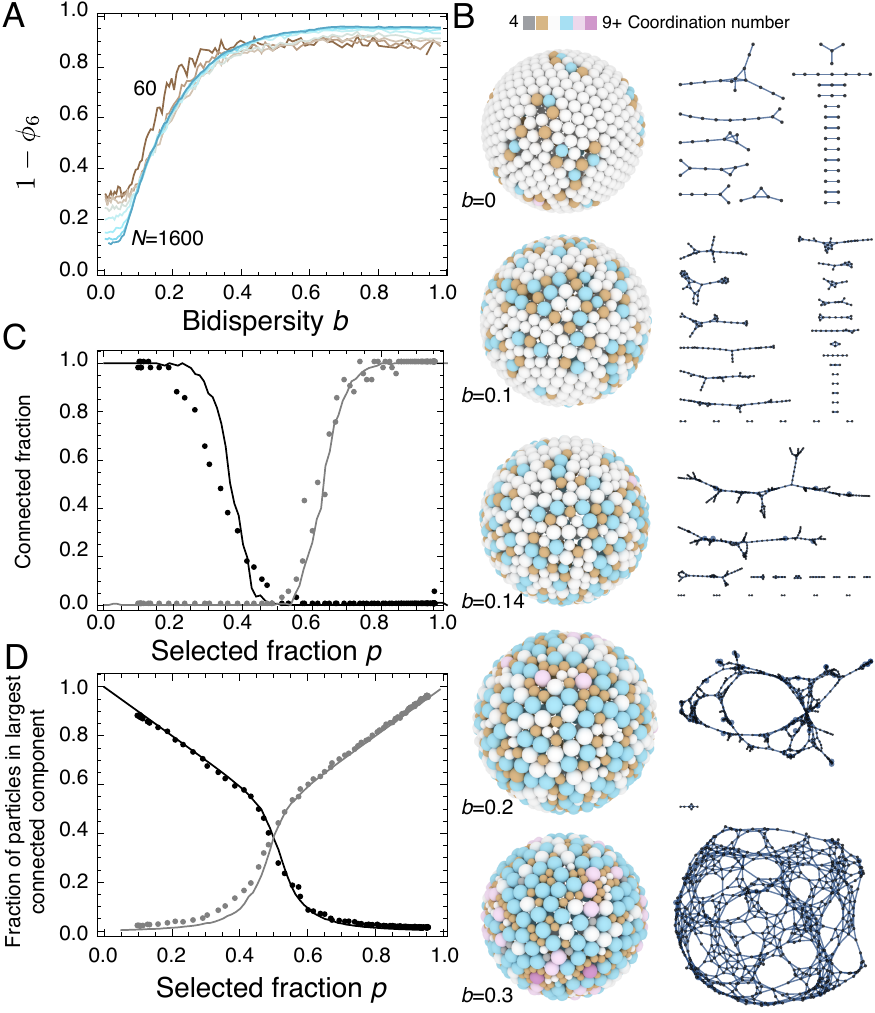}

\caption{\label{fig:Percolation}(Color online) \textbf{A} Fraction of particles
$1-\phi_{6}$ with coordination number $C\protect\neq6$ as a function
of bidispersity. \textbf{B} Representative defect subgraphs different
$b$ illustrating growth and connection of the scar network. \textbf{C}
Comparison with random percolation: a fraction $p$ sites are randomly
selected on a $b=0$, $N=800$ neighbor graph. Shown is the fraction
of simulations where the selected sites form a connected structure
(gray solid line) and the fraction where the non-selected sites retain
global connectivity (black solid line). Points show the fraction of
simulations where the hexatic (black points) and non-hexatic subgraphs
remain connected. \textbf{D} Size of the largest connected component
for random percolation (solid lines) and bidispersed neighbor graphs
(points). }
\end{figure}

We now examine the coordination number directly.  In fig. \ref{fig:OrderParameter}C,
we plot the average coordination number per particle, separated into
large-large, large-small and small-small contacts and for different
$N$. At infinitesimal $b$, each particle has six neighbors, three
smaller and three larger on average. With increasing $b$, the number
of large-small contacts per particle remains a constant value of three;
larger particles gain more large neighbors while smaller particles
lose small contacts. At the Apollonian point, the smaller particles
are surrounded by three larger neighbors, while the larger particles
are on average surrounded by six large neighbors and three smaller
neighbors. For $b>b_{A}$, the coordination numbers remain constant,
consistent with the discussion above where smaller particles are caged
within the interstices of the larger particles. Smaller values of
$N$ follow similar trends, but tend to have lower coordination numbers.

Finally, we calculated the pair correlation function $g(s)$ that
encodes particle's local environment; results are displayed in Fig.
\ref{fig:OrderParameter}D. For $b=0$, we see persistent peaks at
large $s$ indicative of long range order and a split second peak
in agreement with previous studies in flat space\cite{Donev2005}.
Increasing bidispersity slightly to $b=0.06$ causes the split peak
to disappear, representing the disruption of local crystalline packing,
but the long range order persists. Proceeding to $b=0.13$, $g(s)$
is now flat, indicating that the long range order has disappeared.
This is our first indication that the minimum in $\Phi$ at $b_{c}$
observed in Fig. \ref{fig:Packing} is associated with a transition
where long range crystalline order is disrupted. 

One measure of the abundance of crystallinity is the fraction $\phi_{6}$
that possess a coordination number of $6$. In fig. \ref{fig:Percolation}A,
we plot $1-\phi_{6}$ as a function of bidispersity revealing a transition:
as $b$ increases from zero, $1-\phi_{6}$ is approximately constant
then rises rapidly to unity, reaching a value of $\frac{1}{2}$ at
$b=b_{p}\approx0.15$. Above bidispersity $b\approx0.5$, a vanishing
fraction of particles possess six neighbors. These trends persist
for all values of $N$ shown, but $1-\phi_{6}$ is larger at $b=0$
for small $N$ since topology mandates a minimal number of defects. 

To understand this transition further, it is necessary to examine
the microstructural information encoded in the neighbor graphs, the
adjoint graph of the radical tessellation. We crudely separate the
crystalline and non-crystalline components by deleting from a neighbor
graph all vertices that have six neighbors, yielding the ``non-hexatic''
subgraph. Illustrative examples of these subgraphs are depicted in
fig. \ref{fig:Percolation}B. For $b=0$ the subgraph consists of
small disconnected components corresponding to the previously-studied
scars, which are essentially linear in morphology, with a small number
of branches. As bidispersity increases to $b=0.1$, just below $b_{p}$,
the connected subgraphs are still recognizably scar-like in nature,
but have a branching morphology and are substantially longer. By $b=0.14$,
close to $b_{p}$, the defect subgraph remains disconnected, but is
now dominated by a few large connected graphs that are mostly linear
with branches. Finally, above $b_{p}$ at $b=0.2$ the defect subgraph
is now mostly a single connected structure with a small number of
additional isolated defects; it is no longer branching, but with linear
sections that link into a foam-like structure. For $b=0.3$, the defect
subgraph retains this structure, but is more densely connected. 

The gradual growth and long-range connection of the non-hexatic subgraph
due to bidispersity is therefore a percolation transition that occurs:
As $b$ increases around $b_{p}$, the number of sites participating
in the non-hexatic subgraph increases until they form a connected
structure. Percolation transitions are well-studied\cite{grimmett1997percolation}.
The canonical formulation is: given a network, and selecting a fraction
$p$ sites, what is the probability that one of the selected sites
belongs to a long-range connected structure? Clearly, the system under
consideration cannot be precisely mapped onto this problem because
the neighbor graph changes with $b$. However, by averaging over all
particle pairs in Fig. \ref{fig:OrderParameter}B we see that the
mean coordination number remains 6 for all $b$. Thus, we examine
the canonical percolation problem on the neighbor graph of a monodisperse
packing for $N=800$ particles. From such a graph, we randomly select
a fraction $p$ sites and repeat this procedure to form $n$ trials.
Plotted in Fig. \ref{fig:Percolation}C is the fraction of trials
where the selected components form a connected structure (gray line)
and where the remaining components retain their connectivity (black
line). We compare this to the bidispersity percolation transition
by the placing the non-hexatic subgraph in correspondence to the selected
subgraph in the random percolation model; the selected fraction is
therefore $p=1-\phi_{6}$. The fraction of connected hexatic and non-hexatic
subgraphs at each value of $p$ is plotted as points in Fig. \ref{fig:Percolation}C,
showing that the percolation thresholds are in good agreement. Notably,
the hexatic subgraph become disconnected around $p\approx0.4$, which
occurs at $b\lesssim b_{p}=0.15$ in Fig. \ref{fig:OrderParameter}A.
Percolation implies a growing lengthscale, so we also computed the
size of the largest connected component of the selected and unselected
subgraphs, plotted in solid lines in Fig. \ref{fig:Percolation}D.
Again calculating corresponding values from the bidispersed neighbor
graphs, shown as points in Fig. \ref{fig:Percolation}D, we see excellent
agreement. We infer from this that the qualitative features of the
bidispersity percolation transition are predicted by a random percolation
transition on the monodisperse neighbor graph. 

\begin{figure}
\includegraphics[width=1\columnwidth]{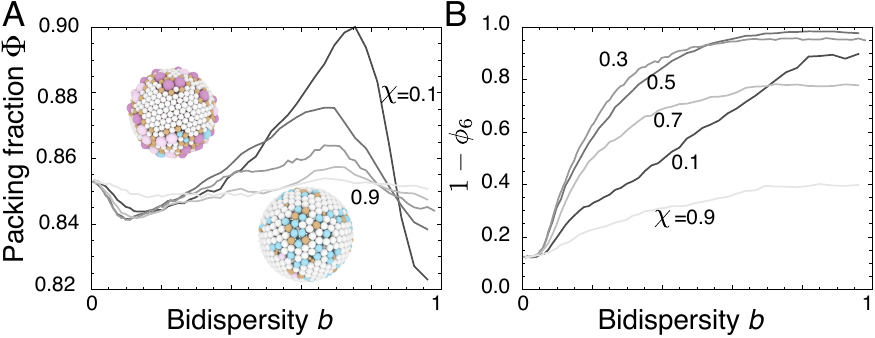}

\caption{\label{fig:Intermix}\textbf{A }Packing fraction and \textbf{B} fraction
of particles $1-\phi_{6}$ with coordination number $C\protect\neq6$
as a function of bidispersity for different number fractions of large
particles $\chi=N_{L}/N$. \emph{Insets:} representative packings
shown for $b=0.33$ and $\chi=0.1$ (upper) and $\chi=0.9$ (lower). }
\end{figure}

To test this, we attempted to disrupt the transition by varying the
fraction of large particles $\chi=N_{L}/N$, motivated by the idea
that growth of the scars might be prevented if sufficiently few minority
particles are present. The packing fraction for several values of
$\chi$ is shown in fig. \ref{fig:Intermix}A. Small $\chi$ leads
to a dramatic enhancement of the packing fraction at the Apollonian
point, but $\chi=0.9$ flattens it as well as suppresses the low $b$
minimum. Looking at the defect subgraphs, those in $\chi=0.9$ do
not exhibit connected defect subgraphs. For a given $\chi$, bidispersity
determines $1-\phi_{6}$, which is the parameter that determines whether
we have percolation or not: Examining this, plotted in fig. \ref{fig:Intermix}B,
shows that for $\chi=0.9$, $1-\phi_{6}$ is just short of the threshold
$\sim0.4$ for random percolation. 

\emph{Conclusion\textemdash }We have shown that the packing fraction
of bidispersed packings of spheres on a spherical surface is determined
by three influences: an Apollonian packing for $b\approx0.73$ where
small particles fit into the interstices of large particles produces
a global maximum; commensurate eight and tenfold coordinated configurations
of particles yield an inflexion point at $b\approx0.41$; a minimum
at $b\approx0.1$ is due to the growth and percolation of ``scars''
previously observed in the monodispersed case. By adjusting the ratio
of large particles, we have shown that preventing the percolation
transition greatly attenuates the minimum. The growing lengthscale
and critical fraction necessary for percolation were found to be in
agreement with those for random percolation on the monodispersed neighbor
graph. 

\emph{Acknowledgement\textemdash The authors thank the Research Corporation
for Science Advancement for funding through a Cottrell Award. }

\end{document}